\documentclass[11pt,a4paper]{article}
\usepackage[T1]{fontenc}
\usepackage[utf8]{inputenc}
\usepackage{authblk}

\usepackage{amssymb,amsmath,graphicx,graphics,color}

\newcommand{\ket}[1]{|#1\rangle}

\newcommand{\im}[1]{\textrm{Im}\left\{  #1\right\}}

\newcommand{\de}{\partial}
\newcommand{\sech}[1]{\textrm{sech}\left(  #1\right)}
\newcommand{\eq}[2]{\begin{equation} \label{#1} #2 \end{equation}}
\newcommand{\eps}{\epsilon}

\newcommand{\etal}{{\em et al.}}

\newcommand{\kk}{\mathbf{k}}
\newcommand{\pp}{\mathbf{p}}

\newcommand{\EE}{\mathbf{E}}
\newcommand{\JJ}{\mathbf{J}}

\newcommand{\rr}{\mathbf{r}}

\newcommand{\AAA}{\mathbf{A}}

\newcommand{\sss}{\boldsymbol{\sigma}}

\begin{document}

\title{\bf Dynamical centrosymmetry breaking in graphene}
\author{David N. Carvalho$^1$, Andrea Marini$^2$, Fabio Biancalana$^1$\\
$^1$School of Engineering and Physical Sciences, Heriot-Watt University, EH14 4AS Edinburgh, UK\\
$^2$ICFO-Institut de Ciencies Fotoniques, 08860 Castelldefels (Barcelona), Spain}

\maketitle

\begin{abstract} We discover an unusual phenomenon that occurs when a graphene monolayer is illuminated by a short and intense pulse at normal incidence. Due to the pulse-induced oscillations of the Dirac cones, a dynamical breaking of the layer's centrosymmetry takes place, leading to the generation of second harmonic waves. We prove that this result can only be found by using the full Dirac equation and show that the widely used semiconductor Bloch equations fail to reproduce this and some other important physics of graphene. Our results open new windows in the understanding of nonlinear light-matter interactions in a wide variety of new 2D materials with a gapped or ungapped Dirac-like dispersion. 
\end{abstract}

\paragraph{Introduction ---} The physics of graphene and related materials has attracted a broad interest since the initial experimental realization of graphene monolayers \cite{graph1}. At relative low energies, graphene shows a unique Dirac-like band structure and this implies that quasielectrons behave as if they were massless Dirac fermions \cite{Novoselov2005}. Due to this special property, graphene electronics is quite different from conventional semiconductor electronics, and holds the promise of revolutionising the technological landscape in many different ways \cite{Novoselov2005}. Apart from its noteworthy electronic properties, graphene also shows extraordinary optical properties \cite{Bonaccorso2010} which have already been employed in photonics for ultrafast photodetectors \cite{Mueller2010}, optical modulation \cite{Liu2011}, 
molecular sensing \cite{Marini2015bis}, and several nonlinear applications \cite{Gullans2013,Smirnova2014}. Graphene's optical response is characterized by a highly-saturated absorption at rather modest light intensities \cite{Bao2009}, a remarkable property which has already been exploited for mode-locking in ultrafast fiber-lasers \cite{Sun2010}. The high nonlinear response of graphene leads to the efficient generation of higher harmonics \cite{Peres2014,Mikhailov2014}. Theoretical approaches to model the nonlinear dynamics of graphene 
typically rely on the Boltzmann transport equation, accounting only for intraband electron dynamics \cite{Mikhailov2007}, and on the popular semiconductor Bloch equations (SBEs), which account only for the interband dynamics adapted to the conical dispersion \cite{malic}.

In this paper we show the existence of a previously unknown nonlinear optical phenomenon that occurs when a graphene monolayer is illuminated by a short and intense pulse at normal incidence. Due to the pulse-induced oscillations of the Dirac cone, a dynamical breaking of the layer's centrosymmetry takes place, leading to the generation of second harmonics. The importance of this novel nonlinear effect is that it can only be found by using the full Dirac equation, while the SBEs completely fail to describe it, and we explain the deep motivations behind this failure in the latter equations.

\begin{figure}
\centering
\includegraphics[width=10cm]{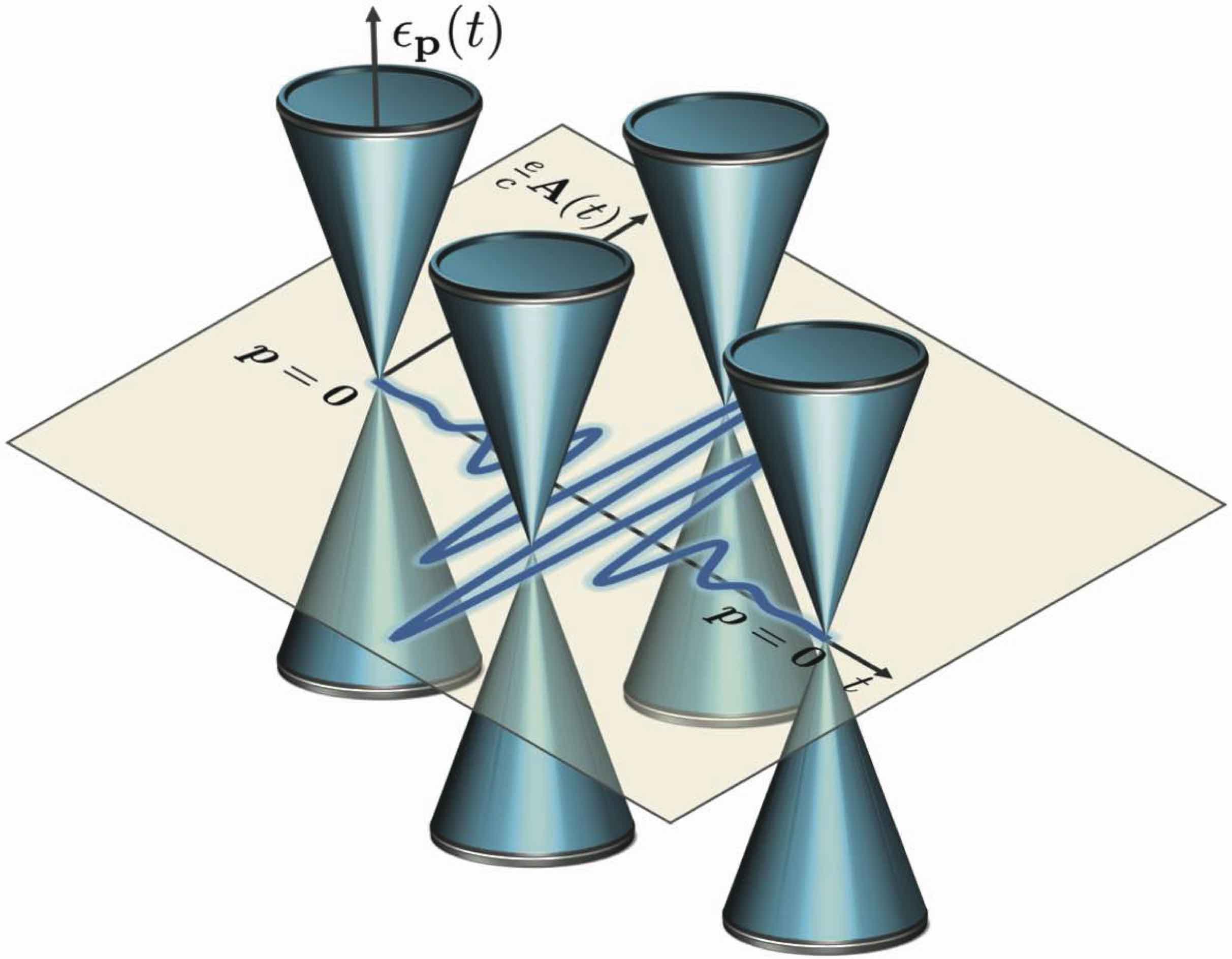}
\caption{(Color online) Sketch showing the dynamical centrosymmetry breaking mechanism in graphene, when illuminated by short, intense pulses at normal incidence. The Dirac cone is shaken from the $p_{x}=p_{y}=0$ position by the time-dependent pulse momentum $\frac{e}{c}A(t)$, which without loss of generality acts only along the $p_{x}$ direction (the electric field is linearly polarised for simplicity). This motion continuously break the $\kk\rightarrow-\kk$ symmetry in momentum space, and therefore also the $\rr\rightarrow-\rr$ symmetry in direct space, inducing a breaking of centrosymmetry of the graphene layer. Crucially, this dynamical centrosymmetry breaking produces SHG waves that do not average out to zero after the integration over the momenta, as explained in the main text.}
\label{fig1}
\end{figure}

\paragraph{Governing equations ---} 
The starting point of this discussion is the Dirac equation for a massless electron in graphene, interacting with an electromagnetic vector potential $\AAA$:
\eq{dirac1}{i\hbar\de_{t}\ket{\psi_{\kk}(t)}=v_{\rm F}~\sss\cdot\left(\pp+\frac{e}{c}\AAA\right)\ket{\psi_{\kk}(t)},}
where $v_{\rm F}$ is the Fermi velocity in a graphene monolayer, $c$ is the speed of light in vacuum, $-e$ is the electron charge, $\sss\equiv(\sigma_{x},\sigma_{y})$ is the 2D Pauli-matrix vector, and $\ket{\psi_{\kk}(t)}$ is the time-dependent 2-spinor representing electrons in the conduction and valence bands for a specific electronic momentum $\pp \equiv \hbar\kk$. We work at normal incidence conditions, in the Coulomb gauge $\nabla \cdot \AAA = 0$ and without loss of generality we assume that the electric field $\EE \equiv -(1/c)\de_{t}\AAA$ is linearly polarised along the (arbitrary) $\hat{x}$ axis. 

General analytical solutions of Eq.~(\ref{dirac1}) are not known. However, following a procedure originally outlined in two seminal papers by Ishikawa \cite{ishikawa1,ishikawa2}, one can seek solutions using band eigenfunctions: $\ket{\psi_{\kk}(t)} = c_{+}(t)\ket{\psi_{+1, \kk}(t)} + c_{-}(t)\ket{\psi_{-1,\kk}(t)}$, where $\ket{\psi_{\lambda, \kk}(t)} \equiv (e^{-\frac{i}{2}\theta_{\kk}(t)},\lambda e^{\frac{i}{2}\theta_{\kk}(t)})e^{-i\lambda \Omega_{\kk}(t)}/\sqrt{2}$, and $\lambda=+1$ ($\lambda=-1$) refers to to conduction (valence) band. In this fashion, the field interaction is accounted for by generalising the field-free electron eigenstates to a time-dependent {\em ansatz}. Here, $\theta_{\kk}(t) \equiv \arctan(p_{y}/[p_{x} + \frac{e}{c}A(t)])$ is the dynamical angle and $\Omega_{\kk}(t) \equiv(v_{\rm F}/ \hbar)\int_{-\infty}^{t}[(p_{x}+\frac{e}{c}A(t'))^{2}+p_{y}^{2}]^{1/2}dt'$ is the dynamical phase with the instantaneous energy, both corrected by the photon momentum \cite{ishikawa1,ishikawa2}.

By writing the time derivatives of $c_{\pm}$, and introducing the new dynamical variables `population' $w_{\kk} \equiv |c_{+}|^{2}-|c_{-}|^{2}$ and `polarisation' $q_{\kk} \equiv c_{+}c_{-}^{*}e^{-2i\Omega + i\omega_{0}t}$, where $\omega_{0}$ is the central frequency of the input pulse, one can show that the full Dirac equation (\ref{dirac1}) can be rewritten (without using any approximations) in the following form, which we call the {\em Dirac-Bloch equations} (DBEs):
\begin{eqnarray}
\dot{q}_{\kk} + i(2 \dot{\Omega}_{\kk} - \omega_{0} - i\gamma_{2})q_{\kk} + \frac{i}{2} w_{\kk} \dot{\theta}_{\kk} e^{i\omega_{0}t} = 0, \label{db1} \\
\dot{w}_{\kk} + \gamma_{1}(w_{\kk} - w^{0}_{\kk}) + i \dot{\theta}_{\kk} \left(q_{\kk} e^{-i\omega_{0}t} - q^{*}_{\kk} e^{i\omega_{0}t}\right) = 0,\label{db2} 
\end{eqnarray}
where $\dot{\Omega}_{\kk}\equiv(v_{\rm F}/\hbar)[(p_{x}+\frac{e}{c}A(t))^{2}+p_{y}^{2}]^{1/2}$ and $\dot{\theta}_{\kk}\equiv ep_{y}E(t)/[(p_{x} + \frac{e}{c}A(t))^{2}+p_{y}^{2}]$. Dephasing effects are included phenomenologically through the coefficients $\gamma_{1,2} \equiv1/T_{1,2}$ whereas the effect of intrinsic, field-independent parameters such as temperature and chemical potential can be incorporated in a momentum-dependent equilibrium value of the populations, $w^{0}_{\kk}$. Typically, at zero temperature one has $w^{0}_{\kk} = -1$, implying that all carriers are initially in the valence band, irrespective of their momenta. For non-vanishing temperatures and/or a non-vanishing chemical potential $\mu$, the starting population is given by $w^{0}_{\kk} = -\sinh(x)/(\cosh(x)+\cosh(y))$, where $x\equiv\hbar|\kk|v_{\rm F}/(k_{\rm B}T)$ and $y\equiv\mu/(k_{\rm B}T)$.
In the absence of dephasing ($\gamma_{1}=\gamma_{2}=0$) the law of conservation of probability for each `two-level system' of wavevector $\kk$ ($w_{\kk}^{2}+4|q_{\kk}|^{2}=1$) is also satisfied. This is approximately true for ultrashort pulses in the coherent regime, i.e. for pulse durations much shorter than the dephasing times, $t_{0}\ll T_{1,2}$, where $t_{0}$ is the input pulse duration. In this paper we assume for simplicity that we are in such regime. However, our results are very general and also valid in the presence of dephasing and are in no way limited by the probability conservation.

Equations (\ref{db1}-\ref{db2}) must be directly compared with the well-known {\em semiconductor Bloch equations} (SBEs), universally used in the theoretical description of gapped semiconductors \cite{lindberg,haug} and also largely employed in simulations of light interaction with graphene \cite{malic,knorr1,stroucken,voss1}. In order to reduce Eqs. (\ref{db1}-\ref{db2}) to the SBEs, one must neglect the contribution of the photon momentum $eA(t)/c$ in the quantities $\dot{\Omega}_{\kk}$ and $\dot{\theta}_{\kk}$, obtaining exactly the SBEs used, for instance, in Ref. \cite{malic}. That this reduction is {\em never} acceptable in gapless media will be proved shortly. It follows that despite their popularity, the use of SBEs for graphene is inadequate when dealing with pulses that are short and/or intense enough to shift considerably the position of the Dirac point in momentum space.

Another important assumption that is implicitly made in the derivation of Eqs. (\ref{db1}-\ref{db2}) is the {\em absence of effective Coulomb interactions amongst the carriers}. The effect of the Coulomb interaction is usually implemented (for instance in the theory of conventional semiconductors) by `renormalising' the Rabi frequency and the band energies by coupling all the momenta $\kk$ of the band structure, see Ref. \cite{haug} for a complete review on the subject. However, it is surprising that in graphene such renormalisation must be absent or very small. We refer the reader to the textbook by Katsnelson (see Ref. \cite{katsnelson}, p. 167) and to several recent works on the subject \cite{yang,mishchenko,giuliani,dassarma}.


One can make two crucial observations when looking at Eqs. (\ref{db1}-\ref{db2}). Firstly, the dipole moment that multiplies the electric field inside the function $\dot{\theta}_{\kk}(t)$ is given by $M_{\kk}\equiv \frac{\hbar}{2}ep_{y}/[(p_{x}+\frac{e}{c}A(t))^{2}+p_{y}^{2}]$, and it is therefore a {\em time dependent} quantity. This is very unusual in the theory of two-level systems and we are not aware of any other realistic physical situations in which the dipole moment is temporally oscillating with the pulse.  Secondly, the frequency detuning between a specific two-level system with wavevector $\kk$ and the pulse frequency $\omega_{0}$ is also oscillating in time as $2\dot{\Omega}_{\kk}-\omega_{0}=\frac{2v_{\rm F}}{\hbar}[(p_{x}+\frac{e}{c}A(t))^{2}+p_{y}^{2}]^{1/2}-\omega_{0}$ [see second term in Eq.~(\ref{db1})]. In other words, the pulse itself modulates the band structure continuously, leading to {\em global dynamical oscillations} of the Dirac cone. We shall see the significance of such a modulation in the generation of new harmonics when discussing the centrosymmetry breaking effect below. These features are not present in the SBEs, since the photon momentum is neglected in that formulation.

Once Eqs. (\ref{db1}-\ref{db2}) are solved numerically for each value of $\kk$, it is possible to find the integrated current vector, which takes into account both electron and hole contributions:
\eq{current1}{\JJ = -ev_{\rm F}\frac{g_{\rm s}g_{\rm v}}{d(2\pi)^{2}}\int
\left(\begin{array}{cc} \cos{\theta}_{\kk} & \sin{\theta}_{\kk} \\ \sin{\theta}_{\kk} & -\cos{\theta}_{\kk} \end{array}\right)
\left(\begin{array}{c} w_{\kk} - w^{0}_{\kk} \\ -2\im{q_{\kk}e^{-i\omega_{0}t}} \end{array}\right)
 d\kk,}
where $g_{\rm s}=g_{\rm v}=2$ are respectively the spin and valley degeneracies, $d \kk \equiv k\ \!\!\!dk\ \!\!\!d\phi$ is the 2D differential in momentum space and $ d = 0.33$ nm is the layer's thickness. We also define the two individual components of the current as $\JJ \equiv(J_{x},J_{y})$. Note that, since the electric field is polarised along the arbitrary direction $\hat{x}$, the integrated current component $J_{y}$ vanishes identically. For the $J_{x}$ component, one can also distinguish between the {\em intraband} current (responsible of electronic transitions within the same band), proportional to $w_{\kk} - w^{0}_{\kk}$ and the {\em interband} current (responsible for vertical transitions between the valence and conduction bands), proportional to $-2\im{q_{\kk}e^{-i\omega_{0}t}}$, in Eq.~(\ref{current1}) - hence $J_{x} = J_{\rm intra} + J_{\rm inter}$. We choose realistic, zero-averaged localised electric and vector potential fields, in order not to introduce unphysical static electric fields: $A(t) = A_{0}\sech{t/t_{0}}\sin(\omega_{0}t)$ and $E(t)=-\de_{t}A/c$.

\paragraph{Dynamical centrosymmetry breaking and SHG generation ---}  The instantaneous energy eigenvalues derived from Eq.~(\ref{dirac1}) are given by $\epsilon_{\lambda, \mathbf{p}}(t) = \lambda v_{\rm F}\sqrt{(p_{x}+\frac{e}{c}A(t))^2+p_{y}^2}$. For $A(t) = 0$, one recovers the unperturbed Dirac cone band structure $\epsilon_{\lambda, \mathbf{p}} = \lambda v_{\rm F}|\pp|$. However, for $A(t) \neq 0$, the whole Dirac cone oscillates around the $\kk = 0$ point together with the pulse along the $p_{x}$ direction due to the pulse polarisation along $\hat{x}$. A graphical depiction of this oscillation is shown in the sketch of Fig. \ref{fig1}. 

During those moments when the Dirac cone is displaced the inversion symmetry in momentum space, namely $\kk\rightarrow -\kk$, is temporarily broken, and so is the inversion symmetry in real space $\rr\rightarrow-\rr$, leading to a {\em dynamical breaking of the centrosymmetry, induced by the pulse itself}. If the pulse is intense enough, this leads to the possibility of radiating SHG radiation at normal incidence, a situation which is normally forbidden due to the supposed centrosymmetry of the graphene lattice. This is a previously unknown effect in graphene and is the central result of this paper. Note that this novel effect must be distinguished from the well-known photon-drag effect - a relativistic phenomenon in which the pulse is illuminated obliquely  on the graphene layer. An oblique angle of incidence induces a fixed longitudinal momentum that permanently displaces the upper and the lower cones of the Dirac dispersion, also leading to experimentally verified THz and SHG effects \cite{drag1}.

Even though the dynamical centrosymmetry breaking effect occurs for every specific value of $\kk$ in the DBEs (\ref{db1}-\ref{db2}), it is far from obvious that this can give non-vanishing SHG radiation after integrating over all momenta. In fact, as we shall see in the next section, even the conventional SBEs have the potential to show SHG radiation emission, due to the fact that $w_{\kk}$ in Eq. (\ref{db2}) has terms oscillating with frequencies zero and $2\omega_{0}$ -- however after integration over all momenta these emissions cancel out exactly to zero in the SBEs case, unlike in the DBEs case.

\paragraph{Simulations ---} We have solved numerically the DBEs [Eqs. (\ref{db1}-\ref{db2})] with an explicit, highly accurate, adaptive sixth-order Runge-Kutta algorithm in which the tolerance of the adaptive scheme is checked by tracking the probability conservation for each value of time and $\kk$. 

\begin{figure}
\centering
\includegraphics[width=10cm]{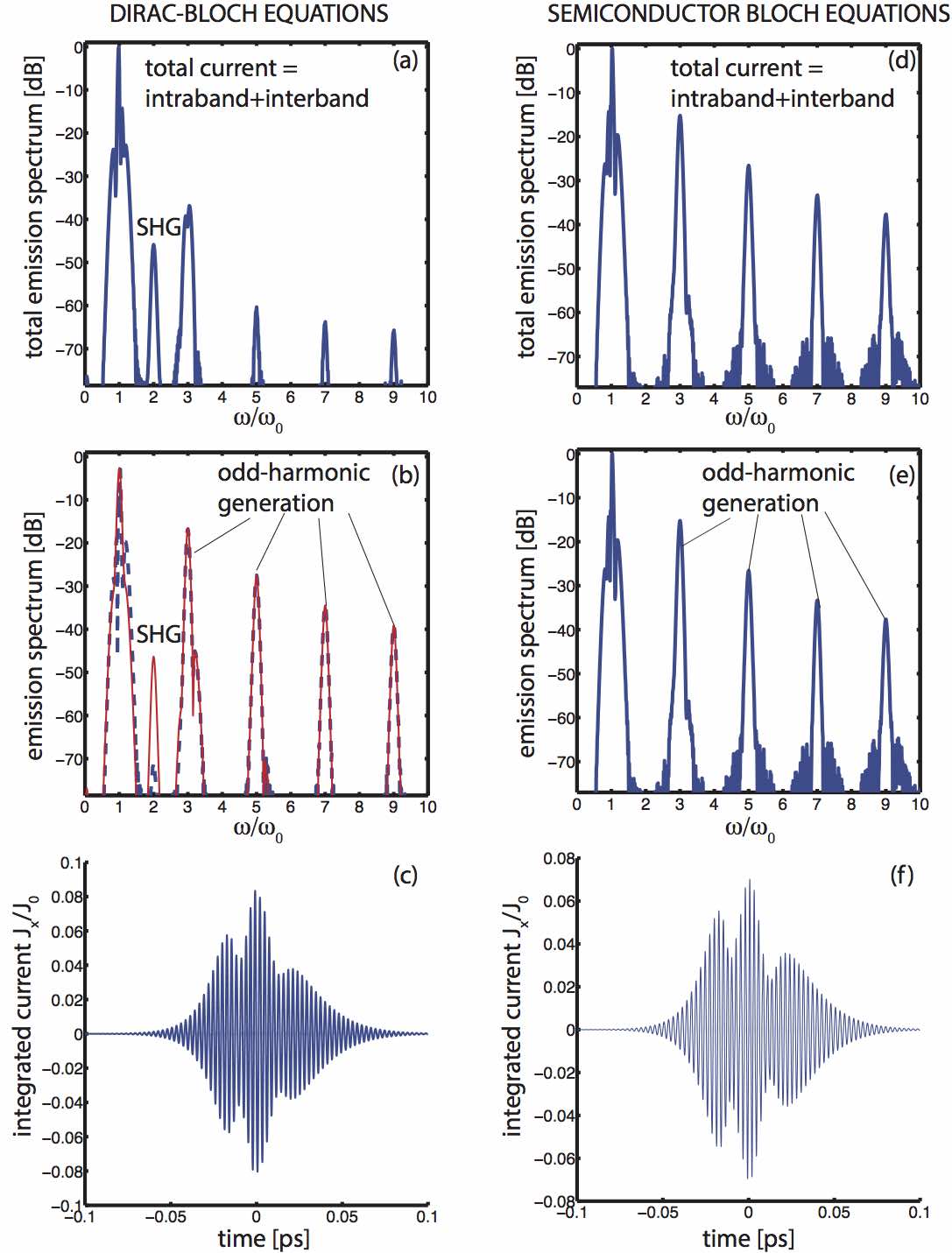}
\caption{(Color online) Total and partial emission spectra and currents excited by a $t_{0}=10$ fsec pulse, normally incident and linearly polarized,
with an input energy $I=114$ GW/cm$^{2}$, at zero temperature. First column (a,b,c) shows results obtained by solving numerically the DBEs (\ref{db1}-\ref{db2}), (a) the total (intraband + interband) spectrum, (b) the separated interband (blue line) and intraband (red line) spectra, and (c) the current $J_{x}(t)$ circulating on the graphene layer  [$J_{0}\equiv-e\omega_{0}^{2}/(4dv_{\rm F})$ is a reference scale]. Similar figures are calculated in (d,e,f) using the SBEs, i.e. neglecting the photon momentum in the DBEs. Note in (b) that the SHG mainly originates from the intraband current, while intraband and interband currents contribute equally to the generation of the odd harmonics. Also, (d) and (e) are identical since in the SBEs the intraband current integrates out to zero, and therefore the SHG cannot be predicted by the SBEs.}
\label{fig2}
\end{figure}

In Fig. \ref{fig2}(a,b,c) (first column) we show the output spectra $S(\omega) = |\omega \JJ(\omega)|^{2}$ in dB and the integrated circulating current $J_{x}(t)$, when pumping the graphene layer with a normally incident pulse, $t_{0}=10$ fsec, central wavelength $\lambda_{0}=800$ nm,  intensity $I=114$ GW/cm$^{2}$, and at temperature $T = 0$ $^{\circ}$K. In Fig. \ref{fig2}(a) we show the output spectrum in dB of the total current, intraband plus interband, when using the DBEs (\ref{db1}-\ref{db2}). The dynamical centrosymmetry breaking described above leads to a relatively strong SHG signal, indicated in the figure, at $\omega/\omega_{0}=2$. This SHG signal is an absolute novelty in the theory of graphene, since it was previously thought to be impossible to obtain such signal in normal incidence conditions. Figure \ref{fig2}(a) also shows the more conventional high-harmonic generation typical of a $\chi^{(3)}$ material, with peaks emitted at odd integer values of $\omega/\omega_{0}$.

Figure \ref{fig2}(b) is the same as Fig. \ref{fig2}(a), but we separate the contributions of the intraband (red solid line) and the interband (dashed blue line) currents. We can notice that for the odd-order high-harmonic generation peaks the intraband and interband currents basically contribute equally to the emission (the two lines are almost superimposed), while the SHG peak mainly comes from the contribution of the intraband current.

Figure \ref{fig2}(c) shows the temporal evolution of the total current $J_{x}(t)$. Note that $J_{y}(t)=0$ after integration over the momenta, as it should be since the pulse is linearly polarised along the $\hat{x}$-axis. However, nothing prevents the study of arbitrary polarisation configurations of the system, yielding in general  $J_{y}\neq 0$.

Figures \ref{fig2}(d,e,f) show the same quantities as (a,b,c), this time calculated by using the SBEs. We can see that no SHG is predicted by the SBEs and therefore these equations ultimately fail to show evidence of important physics contained in the Dirac equation (\ref{dirac1}). One can also notice that Fig. \ref{fig2}(d) and (e) are identical, since the intraband contributions are totally absent in the SBE formulation. Furthermore, the intensities of the odd-harmonic peaks in the SBEs in Fig. \ref{fig2}(d,e) are overestimated with respect to their counterparts obtained by using the DBEs in Fig. \ref{fig2}(a,b).

\begin{figure}
\centering
\includegraphics[width=8cm]{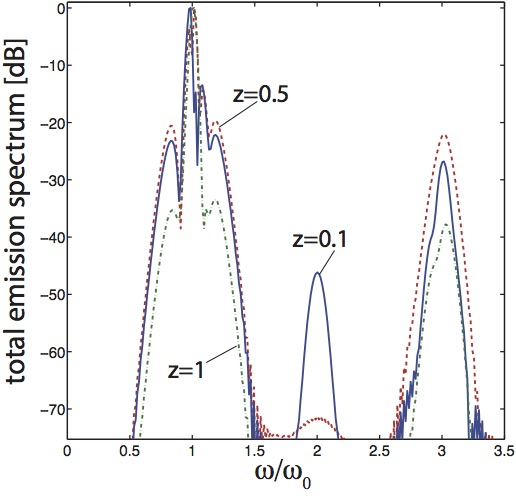}
\caption{(Color online) Total emission spectra for non-vanishing chemical potential $\mu$, parameterised by $z\equiv2\mu/(\hbar\omega_{0})$, for $z=0.1$ (solid blue line), $0.5$ (dashed red line) and $1$ (dashed-dotted green line). Temperature is $T=300$ $^{\circ}$K. One can observe a drastic reduction of the SHG peak when progressively increasing the doping.}
\label{fig3}
\end{figure}

We have also studied the influence of a non-vanishing chemical potential $\mu$ on the output spectra, when using the full DBEs, see Fig. \ref{fig3}. We introduce a scaled parameter $z\equiv2\mu/(\hbar\omega_{0})$ which measures the level of doping relative to half of the input photon energy. When increasing the chemical potential from relatively small values ($z=0.1$) to large values ($z=1$), one can see that the SHG peak rapidly disappears. This is due to the fact that the interband transitions become progressively suppressed when increasing $z$. This leads to a suppression of the intraband transitions too, since the latter depend on the factor $w_{\kk} - w^{0}_{\kk}$ in Eq.~(\ref{current1}), therefore reducing drastically the effects of the dynamical centrosymmetry breaking.

\paragraph{Shortcomings of SBEs for 2D Dirac media ---} We have already mentioned that starting from Eqs. (\ref{db1}-\ref{db2}) it is possible to derive the so-called semiconductor Bloch equations (SBEs), traditionally used in the study of light propagation in semiconductors, by removing the photon momentum $eA/c$ everywhere in the equations. Since their first invention \cite{lindberg}, the SBEs have been fantastically successful in the description of the dynamics of interband transitions in semiconductors, exciton and exciton-polariton formation and spectra, and the semiconductor laser \cite{haug}. Immediately after the explosion of graphene research in recent years, the SBEs adapted to the graphene dispersion have been routinely applied to study the interaction between linear and nonlinear pulses with massless Dirac electrons \cite{malic,knorr1}.

However, we now prove that {\em the SBEs are often inadequate when studying gapless Dirac media} like graphene and, for pulses that are short or intense enough, they will also fail even in the case of gapped Dirac media such as MoS$_{2}$, WSe$_{2}$ and phosphorene (which at the time of writing are receiving much attention for their potentially game-changing electronic and mechanical properties, see e.g. Ref. \cite{reviewdirac}).

In order to show precise conditions for the failure of SBEs, it is useful to extend the model of Eqs. (\ref{db1}-\ref{db2}) to a gapped layer. In order to do this, we insert a mass term in Eq.~(\ref{dirac1}), proportional to the energy gap $E_{\rm g}$:
\eq{dirac2}{i\hbar\de_{t}\ket{\psi_{\kk}(t)}= \left[v_{\rm F}~\sss\cdot\left(\pp+\frac{e}{c}\AAA\right)+\frac{E_{\rm g}}{2}\sigma_{z}\right]\ket{\psi_{\kk}(t)},} where $\sigma_{z}$ is the diagonal Pauli matrix. The opening of the gap leads to the formation of strongly bound excitons and therefore Coulomb interactions must be included in this case. However, the estimates we give below are independent of the dynamical role of these interactions.

From Eq.~(\ref{dirac2}), one can write straightforwardly the instantaneous energy eigenstates as $\epsilon_{\lambda, \mathbf{p}}(t) = \lambda v_{\rm F} \sqrt{(p_{x}+\frac{e}{c}A(t))^{2}+p_{y}^{2}+(E_{\rm g}/2v_{\rm F})^2}$. In the vicinity of the band gap centre ($p_{x} = p_{y} = 0$), the contribution of the photon momentum can be neglected only for those pulse amplitudes satisfying $|eA/c|\ll E_{g}/2v_{\rm F}$. In this case, the DBEs (\ref{db1}-\ref{db2}) are identical to the SBEs, since in this way one eliminates the time dependence of the frequency detuning and the dipole moment $M_{\kk}$. Therefore, the SBEs are a valid description of light-matter interaction in gapped 2D Dirac media only when the pulse spectrum does not overlap substantially with the Dirac point, or when the intensity of the pulse is not too large with respect to the gap energy.
The above condition for the SBEs to be approximately valid can be translated into a condition for the input pulse intensity: $I\ll I_{\rm cr}$, where $I_{\rm cr}\equiv\frac{1}{8}c\eps_{0}\left(\frac{E_{\rm g}\omega_{0}}{ev_{\rm F}}\right)^{2}$.
If the intensity of the pulse is such that $I\geq I_{\rm cr}$, the SBE description loses its validity. To make things worse, even for low-intensity light, short pulses satisfying the condition $\omega_{0}t_{0}<E_{\rm g}/(4\hbar\omega_{0})$ will overlap too much with the Dirac point, also leading to the breaking of the validity of the SBEs. Therefore the SBEs description of gapped Dirac layers is approximately valid only if pulses are neither too short nor too intense.

It is crucial to observe that for gapless media such as graphene, for which $E_{g}=0$, {\em it is in principle never possible to accurately describe light-matter interactions via the SBE}, since the condition $I\ll I_{\rm cr}$ can never be satisfied. That simply means that there are nonlinear optical effects that the SBEs will never be able to describe -- one of those effects being the dynamical centrosymmetry breaking described in the previous section -- making the use of the full DBEs (\ref{db1}-\ref{db2}) compulsory. This is the final important result of this paper.

\paragraph{Conclusions ---} In conclusion, we discover a novel nonlinear optical effect in graphene where a short and intense pulse can modulate the Dirac cone in time, leading to the temporary breaking of the centrosymmetry. This produces an emission of SHG waves that do not average to zero after momentum integration, and are therefore observable, even at normal incidence. This result contrasts with the long-held belief that such generation ought to be absent. This effect must be distinguished from the photon drag effect, which is present only for non-normal incidence and manifests itself as a static displacement of the Dirac cone. In gapped Dirac media, we have also shown the existence of a critical pulse intensity, above which the SBEs lose their validity and are therefore unable to reproduce the nonlinear optics of 2D media. This critical intensity vanishes for gapless graphene, thus in principle one can never describe the nonlinear optics of graphene with the SBEs with sufficient accuracy -- and indeed the SBEs are not able to predict the onset of the dynamical centrosymmetry breaking. Our results open up new avenues for the accurate description of light-matter interactions in 2D monolayers and are applicable to all materials with a massless or massive Dirac dispersion.

\end{document}